\documentclass[12pt]{article}
\usepackage[a4paper, margin=1.5cm]{geometry}
\usepackage{color}
\usepackage{epsfig}
\usepackage{amsfonts}
\usepackage{mathrsfs}
\usepackage{bm}
\usepackage{array}
\usepackage[title]{appendix}
\usepackage{graphicx}
\usepackage{multirow}
\usepackage{amsmath}
\usepackage{amssymb}
\usepackage{xcolor}
\usepackage{authblk}
\usepackage{cite}
\usepackage[colorlinks,citecolor=blue,linkcolor=blue]{hyperref}

\title{
$R$-matrix type parametrization of the Jost function for extracting the 
resonance parameters from scattering data
}
\author{
P. Vaandrager$^{\,a}$, M.L. Lekala$^{\,a}$
and
S.A. Rakityansky$^{\,b,c}$\\
\parbox{14cm}{
${}^a$ 
{\small Department of Physics, University of South Africa (UNISA), South 
Africa}\\
${}^b$ 
{\small Department of Physics, University of Pretoria, South Africa}\\
${}^c$
{\small Lab of Theoretical Physics, JINR, Dubna, Russia}
}
}
\date{\today}
\begin{document}
\maketitle

\abstract{
A new method is proposed for fitting non-relativistic  
binary-scattering data and for extracting the parameters of possible 
quantum resonances in the compound system that is formed during the collision. 
The method combines the well-known $R$-matrix approach with the analysis based on 
the semi-analytic representation of the Jost functions. It is shown that such a 
combination has the advantages of both these approaches, namely, the number of 
the fitting parameters remains relatively small (as for the $R$-matrix approach) and 
the proper analytic structure of the $S$-matrix is preserved (as for the Jost 
function method). It is also shown that the new formalism, although closely
related to the $R$-matrix method, has the benefit of no dependence on an 
arbitrary channel radius. The efficiency and accuracy of the proposed method 
are tested using a model single-channel potential. Artificial ``experimental'' 
data generated with this potential are fitted, and its known resonances are 
successfully recovered as zeros of the Jost function on the appropriate sheet 
of the Riemann surface of the energy.  
\maketitle
%
%%%%%%%%%%%%%%%%%%%%%%%%%%%%%%%%%%%%%%%%%%%%%%%%%%%%%%%%%%%%%%%%%%%%%%%%%%%%%
\section{Introduction}
%%%%%%%%%%%%%%%%%%%%%%%%%%%%%%%%%%%%%%%%%%%%%%%%%%%%%%%%%%%%%%%%%%%%%%%%%%%%%
In physics the resonance phenomena are associated with oscillations and waves. 
Since the motion of quantum objects is described by waves, it is not surprising 
that resonant states are ubiquitous in atomic and nuclear processes. In the 
collisions of such microscopic objects these resonances usually manifest 
themselves as certain irregularities in the energy dependence of the scattering 
cross-section \cite{Kukulin}. In some rare cases the cross-section has a 
bell-shape maximum around the resonant collision energy, $E_r$. The width, 
$\Gamma$, of this bell at its half-height determines the half-life of the 
resonant state, $T_{1/2}=\hbar\ln2/\Gamma$. In most cases, however, the 
resonance irregularities are more complicated and it is not possible to 
directly deduce $E_r$ and $\Gamma$ from the measured cross-sections.

There are many different approaches to the problem of extracting these 
resonance parameters from the data \cite{Kukulin, Nstar}. Most of them are 
based on the fact that each resonance corresponds to a pole of the $S$-matrix at 
a complex energy (see, for example, Ref.\cite{Taylor})
\begin{equation}
\label{resEnergy}
    E=E_r-\frac{i}{2}\Gamma\ .
\end{equation}
The common attribute for the majority of different approaches is that the data are 
fitted using the $S$-matrix, which can be parametrized in various ways. The poles of the $S$-matrix obtained from the fitting are then sought at complex energies. Not only do the differences among these approaches consist in the various ways of 
parametrizing the $S$-matrix, but also in the ways the $S$-matrix is continued to complex $E$.

The way in which such a continuation is performed plays an important role, because the $S$-matrix is a multivalued function of $E$ and therefore it is defined on a 
complicated Riemann surface of this complex variable. This becomes especially 
important when the Coulomb forces are present, because in such a case the Riemann 
surface has a particularly complicated topology (see, for example,  Ref. 
\cite{Rakityansky2022}). 
The parametrized $S$-matrix may be very good at real energies, but if a chosen 
functional form of this parametrization has inadequate analytic structure, the 
analytic continuation (especially near the branch point, i.e. near the 
threshold) can be inadequate as well.

A possible solution of this continuation-uncertainty problem was suggested in 
Ref. \cite{Rakityansky2013} (see also the book \cite{Rakityansky2022}), where 
semi-analytic expressions of the Jost functions, 
$f_\ell^{\mathrm{(in)}}(E)$ and $f_\ell^{\mathrm{(out)}}(E)$,  were rigorously 
derived for the case of charged particles. The $S$-matrix is just the ratio, 
\begin{equation}
\label{ffratio}
  S(E)=f_\ell^{\mathrm{(out)}}(E)
  \left[f_\ell^{\mathrm{(in)}}(E)\right]^{-1}\ ,
\end{equation}
of the outgoing and incoming Jost functions. In these expressions of the Jost 
functions, all the factors responsible for the multivaluedness and branching 
of the Riemann surface are given explicitly and in an exact way. It was shown 
that the remaining unknown factors are always single-valued and analytic 
functions of $E$. Any reasonable approximation of these factors does not affect 
proper analytic structure of the Jost functions. This fact allows one to 
reliably continue these functions (and thus the $S$-matrix) from the real energy 
axis to any relevant sheet of the Riemann surface (even if the branch-point is 
nearby). In particular, this is a reliable way of locating resonance poles of 
the $S$-matrix (\ref{ffratio}), when it is constructed from the fitting of experimental data.

In Ref. \cite{Rakityansky2013} it was suggested to approximate the 
single-valued factors of the Jost functions by polynomials of the variable 
$(E-E_0)$, with $E_0$ on the real axis and with the coefficients of these  
polynomials being the free parameters that are used to fit experimental data. 
Such an approach was tested on a model problem \cite{Vaandrager2016} and then 
applied to analyse several nuclear reactions  
\cite{Vaandrager2019,He5,Rakityansky2019,Vaandrager2020}, where the resonances 
were located and the corresponding residues of the $S$-matrix and the 
Asymptotic Normalisation Coefficient (ANC) were determined. Although the method 
works and proved to be accurate, it has a significant drawback, namely, that 
the number of the fitting parameters is too large (typically a few dozen).

Such a large number is in a drastic contrast with the number of parameters usually required within the most famous and widely used $R$-matrix approach 
\cite{Wigner1947,Descouvemont2010,Azuma2010,Burke}, where 
it is possible to nicely fit the data with the help of just several parameters.
On the other hand, the $R$-matrix method is not without its own drawbacks. Firstly, everything in this method depends on the choice of a rather 
arbitrary channel radius, at which the inner and outer wave functions match. 
Secondly, the analytic structure of the $S$-matrix obtained from the 
$R$-matrix is obscure, which makes its analytic continuation obscure as well. 

In the present work we combine the advantages of both the $R$-matrix  
and the Jost function theories, namely, a small number of the fitting 
parameters (a feature of the $R$-matrix) with proper analytic structure (a feature of the Jost function method). In order to present the basic idea clearly, as well as to demonstrate the accuracy and efficiency of the proposed method, we consider 
a single-channel problem with a model potential. Exactly calculated cross 
sections for this potential are used as artificial ``experimental'' data 
points, which are fitted using the modified Jost-function method. The new 
method is tested by recovering some of the known resonances for this potential. 

In the next section, the original Jost-function fitting procedure for a 
single-channel problem is described. Then, in Section 
\ref{section_R-matrix}, the relationship between the Jost function and 
$R$-matrix methods is shown. Section \ref{section_alternative_Jost} presents 
the proposed modification of the Jost function fitting procedure. In Section 
\ref{section_example}, a test potential is described. The results from performing the modified Jost-function fitting of the artificial data are then given, i.e. the fitting of the total cross-section generated by the test potential, for $\ell_{\mathrm{max}}\leqslant2$. The conclusion follows.

%%%%%%%%%%%%%%%%%%%%%%%%%%%%%%%%%%%%%%%%%%%%%%%%%%%%%%%%%%%%%%%%%%%%%%%%%%%%%%%
\section{Single-channel Jost-function analysis} 
\label{section_jost}
%%%%%%%%%%%%%%%%%%%%%%%%%%%%%%%%%%%%%%%%%%%%%%%%%%%%%%%%%%%%%%%%%%%%%%%%%%%%%%%
A comprehensive presentation of the Jost function theory can be found in the 
book, Ref. \cite{Rakityansky2022}. For the reader's convenience, we give a very 
concise description of the main equations from that book, which are relevant 
to using the Jost functions for the analysis of a set of single-channel 
scattering data.

The single-channel Jost functions are defined via the regular 
solution, $\phi_\ell(E,r)$, of the radial Schr\"odinger equation, which 
(by definition) behaves like the Riccati-Bessel function near the origin\footnote{It should be noted that this definition remains the same even 
if the Coulomb potential is present. This special regular solution is always 
defined in such a way: that near the origin, it behaves the same as for completely free motion. The effect of the Coulomb forces (if any) manifest in the 
behaviour of the regular solution at large distances. The function $j_\ell(kr)$ 
in Eq. (\ref{reg_sol}) cannot be replaced with $F_\ell(\eta,kr)$ because 
otherwise, for a pure Coulomb problem, the Jost functions would be trivial, i.e. 
equal to one.},
\begin{equation}
\label{reg_sol}
  \phi_\ell(E,r)\mathop{\longrightarrow}\limits_{r\to0}
  j_\ell(kr)\ .
\end{equation}
The Jost functions, $f_\ell^{\mathrm{(in)}}(E)$ and 
$f_\ell^{\mathrm{(out)}}(E)$, are the amplitudes of the incoming and outgoing 
Coulomb-modified spherical waves, $H_\ell^{(\mp)}(\eta,kr)$,
in the asymptotic behaviour of this solution,
\begin{eqnarray}
   \phi_\ell(E,r)\ \mathop{\longrightarrow}\limits_{r\to\infty}
   \ H_\ell^{(-)}(\eta,kr)e^{i\delta_\ell^c}f_\ell^{\mathrm{(in)}}(E)+
     H_\ell^{(+)}(\eta,kr)e^{-i\delta_\ell^c}f_\ell^{\mathrm{(out)}}(E)\ ,
   \label{II.Coulomb.SC.RegSolAss}
\end{eqnarray}
where
\begin{eqnarray}
\label{II.Coulomb.pure.Riccati_Coulomb}
   H_\ell^{(\pm)}(\eta,kr)  =  F_\ell(\eta,kr)\mp iG_\ell(\eta,kr)
   \qquad \qquad
  \nonumber  \\
   \mathop{\longrightarrow}\limits_{r\to\infty}
    \mp i \exp\left\{ \pm  i\left[kr-\frac{\ell\pi}{2}
   -\eta\ln (2kr)+\delta_\ell^c\right]\right\}\ ,
\end{eqnarray}
with $F_\ell$ and $G_\ell$ being the standard regular and irregular Coulomb 
functions. The energy dependent quantity $\delta_\ell^c$ is the pure Coulomb 
phase shift,
\begin{equation}
\label{sigmaL}
     \delta_\ell^c(\eta) = \frac{1}{2i}\ln
     \frac{\Gamma(\ell+1+i\eta)}{\Gamma(\ell+1-i\eta)}\ .
\end{equation}
This and the other functions in the above equations depend on the energy via 
the wave number, $k$, and the Sommerfeld parameter, $\eta$:
\begin{equation}
        k^2  = \frac{2\mu E}{\hbar^2}, 
\qquad \eta = \frac{\mu e^2 Z_1 Z_2}{k \hbar^2},
\end{equation}
where $\mu$ is the reduced mass while $Z_1$ and $Z_2$ are the charges.
 
The Jost functions have the following structure (a detailed derivation can be 
found in Sec. 8.2.6 of the book \cite{Rakityansky2022}):
\begin{eqnarray}
  f_{\ell}^{(\mathrm{in/out})}(E) = 
  e^{\mp i\delta_\ell^c} 
  k^{\ell}
  \Biggl\{
  \frac{k}{D_{\ell}(\eta,k)}   A_{\ell}(E)
  - 
   \left[  M(k)\pm i\right]D_{\ell}(\eta,k)B_{\ell}(E)
     \Biggr\},
  \label{Jost_semi-analytic}
\end{eqnarray}
where
\begin{equation}
\label{Dfunction}
   D_{\ell}(\eta,k) = C_{\ell}(\eta) k^{\ell+1}, \qquad M(k) 
   = \frac{2\eta h(\eta)}{C_{0}^2(\eta)},
\end{equation}
\begin{eqnarray}
h(\eta) = \frac{1}{2} \left[
\psi(1+i\eta) + \psi(1-i\eta)
\right]
-\ln \eta, 
\nonumber
\label{heta}
 \\
\psi(z) = \frac{\Gamma^{\prime}(z)}{\Gamma(z)}, \qquad \qquad
\end{eqnarray}
and the Coulomb barrier factor,
\begin{eqnarray}
  C_\ell(\eta)=
  \frac{e^{-\pi\eta/2}}{\Gamma(\ell+1)}
  \exp \Biggl\{\frac12 \Bigl[\ln\Gamma(\ell+1+i\eta)+
  \ln\Gamma(\ell+1-i\eta) \Bigr] \Biggr\}, \
  \label{II.Coulomb.pure.CoulombC}
\end{eqnarray}
is defined in such a way that for neutral particles it becomes 
unity, $C_\ell(0)=1$, for all $\ell$. This means that the barrier factor 
(\ref{II.Coulomb.pure.CoulombC}) is obtained from the one defined, for example, 
in Ref. \cite{Abramowitz}, by multiplying with $(2\ell+1)!!$.

It should be emphasized that the representation (\ref{Jost_semi-analytic}) is 
exact and valid for any single-channel, two-body system. For any chosen 
two-body potential, all the explicit factors remain the 
same in this representation. The potential only determines the unknown factors $A_\ell(E)$ and  
$B_\ell(E)$. The important property of these unknown factors is that they are 
always single-valued, analytic functions of the variable $E$, and therefore are 
defined on a simple complex-energy plane without any branching points. All the 
troubles associated with the multivaluedness and branching of the Riemann 
surface are correctly taken into account by the explicitly given factors in 
(\ref{Jost_semi-analytic}). Any reasonable approximation of $A_\ell(E)$ and  
$B_\ell(E)$ does not affect the correct analytic structure of the Jost 
functions.

The $S$-matrix (\ref{ffratio}) in the single-channel case takes the form of a 
simple ratio,
\begin{equation} 
\label{S_AB}
        S_{\ell}(E) =  
        e^{2 i\delta_\ell^c} 
        \frac{ k  A_{\ell}(E) 
        - \left[  M(k)- i  \right]D^2_{\ell}(\eta,k)  B_{\ell}(E)}{
        k   A_{\ell}(E) 
        - \left[  M(k)+ i  \right]D^2_{\ell}(\eta,k)  B_{\ell}(E)} .
\end{equation}
Experimental data-points that are fitted via this $S$-matrix, are 
either the phase-shifts or the scattering cross-sections at specific energies.  
The total scattering cross-section,
\begin{equation} 
\label{sigma_total}
\sigma_{\mathrm{total}}(E) = 
\sum_{\ell=0}^{\ell_{\mathrm{max}}} \sigma_{\ell}(E)\ ,
\end{equation}
is the sum of all the partial-wave cross-sections,
\begin{equation} 
\label{sigma_partial}
     \sigma_{\ell}(E) = \frac{\pi}{k^2} (2\ell+1) 
     \vert S_{\ell}(E) - 1\vert^2 ,
\end{equation} 
up to some $\ell_{\mathrm{max}}$ that is deemed to make a noticeable 
contribution.

In the original method based on the semi-analytic representation 
(\ref{Jost_semi-analytic}), the functions $A_\ell(E)$ and  
$B_\ell(E)$ were approximated by several terms of the Taylor expansions around 
a point $E_0$ on the real axis,
\begin{eqnarray}
   A_\ell(E)\approx\sum_{n=0}^{N}a_n(\ell,E_0)(E-E_0)^n\ ,
   B_\ell(E)\approx\sum_{n=0}^{N}b_n(\ell,E_0)(E-E_0)^n\ ,
   \label{orig_Taylor}
\end{eqnarray}
where the coefficients $a_n$ and $b_n$ served as the fitting parameters. 
The only drawback of this method is a large number of the parameters. In order 
to reduce this number, we propose a different approximation of $A_\ell(E)$ and  
$B_\ell(E)$, which is inspired by the $R$-matrix theory. The Jost function 
method modified in this way is described further down in Sec.  
\ref{section_alternative_Jost}.

%%%%%%%%%%%%%%%%%%%%%%%%%%%%%%%%%%%%%%%%%%%%%%%%%%%%%%%%%%%%%%%%%%%%%%%%%%%%%%%%
\section{Relation between the $R$-matrix and the Jost functions} 
\label{section_R-matrix}
%%%%%%%%%%%%%%%%%%%%%%%%%%%%%%%%%%%%%%%%%%%%%%%%%%%%%%%%%%%%%%%%%%%%%%%%%%%%%%%%
The following expression of the single-channel Jost function in terms of the 
$R$-matrix was obtained in Ref. \cite{Vaandrager2024}, 
% where it was used to calculate the Jost 
% function from the computational $R$-matrix.  
% In deriving this expression, the 
% behaviour of the regular radial wave-function near the origin was used. In order 
% to be consistent with \cite{Rakityansky2022}, the regular solution to the radial 
% Schr\"{o}dinger equation is normalised differently to Ref. \cite{Vaandrager2024} 
% and the Jost function is defined slightly differently. This results in an 
% expression for the Jost function which differs from the one given in Ref. 
% \cite{Vaandrager2024} by a finite energy-dependent factor, which (as is known 
% \cite{Rakityansky2022}) does not affect the calculation of physical quantities 
% such as cross-sections or bound and resonance states.   
% 
% Following the same steps as in Ref. \cite{Vaandrager2024}, with the 
% regular solution behaving like the Riccati-bessel function at the 
% origin, it can be shown that the Jost function is related to the $R$-matrix as 
% follows: 
\begin{eqnarray}
\label{FinoutR}
 \nonumber
  f_{\ell}^{(\mathrm{in/out})}(E)
  =
  \frac{\pm ie^{\mp i\delta_\ell^c}k^\ell}{Q_\ell(E,B_R)}
  \Biggl\{\phantom{\frac{\frac{|}{|}}{\frac11}}\!\!\!\!
  H_\ell^{(\pm)}(\eta,ka)\ -   \qquad  \qquad  \qquad   \qquad   \qquad   \qquad  \qquad
  \\[3mm]
  \left[\left.a\frac{dH_\ell^{(\pm)}(\eta,kr)}{dr}\right|_{r=a}
  -B_RH_\ell^{(\pm)}(\eta,ka)\right]
  R_\ell(E,B_R) \Biggr\},
\end{eqnarray}
where $B_R$ is the so-called boundary parameter. Although this parameter is 
present in Eq. (\ref{FinoutR}), it can be shown that the Jost 
function and therefore the $S$-matrix are actually independent of $B_R$ 
\cite{Descouvemont2010,Vaandrager2024}. This is why it is usually chosen as 
$B_R=0$ in the majority of analyses. The $R$-matrix has the following standard 
form \cite{Descouvemont2010, Burke}:
\begin{equation} 
\label{R_matrix}
   R_{\ell} (E,B_R) =  
   \sum_{n=1}^N \frac{[\gamma_{n\ell}(B_R)]^2}{E_{n\ell}-E}\ ,
\end{equation}
where the values $\gamma_{n\ell}$ are either the fitting parameters 
(phenomenological $R$-matrix) or can be calculated with a given potential 
(computational $R$-matrix). 

The parameter, $a$, is very important in the 
$R$-matrix theory. This is the distance at which the ``inner'' and the 
``outer'' wave functions match, i.e. it is assumed that at $r=a$, the wave 
function converges to its asymptotic form. The choice of this parameter is 
rather arbitrary, which is one of the flaws of the $R$-matrix theory. 

The real 
energies $E_{n\ell}$ ($n=1,2,...N$) are the parameters that, in fact, make the 
$R$-matrix theory very efficient. It is with these parameters that one can introduce some prior (approximate) knowledge about the physical 
system into the analysis. From the outset, these energies are fixed to known bound states and to the real parts of known (or approximately known) resonance energies. 

The function, $Q_{\ell}(E,B_R)$, is similar to the $R$-matrix 
\cite{Vaandrager2024}: 
\begin{equation} 
\label{Q_fitting}
  Q_{\ell}(E,B_R) = \sum_{n=1}^{N} 
  \frac{\lambda_{n\ell}(B_R) \gamma_{n\ell}(B_R)}{E_{n\ell}-E}\ ,
\end{equation} 
where $\lambda_{n\ell}(B_R)$ are unknown, real quantities. The function, 
$Q_{\ell}(E,B_R)$, is not required in determining any measurable physical 
values. Indeed, this function is a common factor in (\ref{FinoutR}) for both 
$f_\ell^{\mathrm{(in)}}(E)$ and 
$f_\ell^{\mathrm{(out)}}(E)$ and therefore it cancels out in the $S$-matrix 
(\ref{ffratio}):
\begin{equation}
\label{SR}
  S_\ell=
  -e^{2i\delta_\ell^c}\frac
  {H_\ell^{(-)}-[aH^{\,\prime\,(-)}_\ell-B_RH_\ell^{(-)}]R_\ell}
  {H_\ell^{(+)}-[aH^{\,\prime\,(+)}_\ell-B_RH_\ell^{(+)}]R_\ell}\ ,
\end{equation}
where, in order to simplify notation, we drop all the arguments of the 
functions and denote by a prime the derivative over $r$. In fact, the numerator 
and denominator in this equation are given by Eq. (\ref{FinoutR}) and therefore 
it is easy to recover these arguments, when necessary.

We have two different representations of the Jost functions: first, by Eq. 
(\ref{Jost_semi-analytic}) in terms of $A_\ell(E)$ and $B_\ell(E)$, and 
second, by Eq. (\ref{FinoutR}) in terms of the $R$-matrix. Their equivalence 
can be shown if we choose the functions $A_\ell(E)$ and $B_\ell(E)$  in the 
following special forms:
\begin{eqnarray} 
   A_{\ell}(E) = \frac{D_{\ell}(\eta,k)}{kQ_{\ell}(E,B_R)} 
   \Biggl\{
   J_{\ell}(\eta,ka)
   -  
    \left[
   a\left.
   \frac{dJ_{\ell}(\eta,kr)}{dr}\right|_{r=a}
   -B_R J_{\ell}(\eta,ka)
   \right]R_{\ell}(E,B_R) 
   \Biggr\} 
   \nonumber \\
   \label{A_R}
\end{eqnarray}
and
\begin{eqnarray} 
   B_{\ell}(E) = \frac{\left[D_{\ell}(\eta,k)\right]^{-1}}{Q_{\ell}(E,B_R)} 
   \Biggl\{
   F_{\ell}(\eta,ka)
   -  
    \left[
   a\left.
   \frac{dF_{\ell}(\eta,kr)}{dr}\right|_{r=a}
   -B_R F_{\ell}(\eta,ka)
   \right]R_{\ell}(E,B_R) 
   \Biggr\}\ ,
   \nonumber \\
    \label{B_R}
\end{eqnarray}
where
\begin{equation}
   J_{\ell}(\eta,kr) =  G_{\ell}(\eta,kr) - M(\eta)F_{\ell}(\eta,kr)\ .  
\end{equation}
As we know \cite{Rakityansky2022}, the functions $A_{\ell}(E)$ and 
$B_{\ell}(E)$ in Eq. (\ref{Jost_semi-analytic}) are single-valued and 
analytic. However, when written in terms of the $R$-matrix as given above, 
their analyticity is not obvious. In order to check if Eqs.  
 (\ref{A_R}) and (\ref{B_R}) give single-valued analytic functions, we 
transform them by replacing the Coulomb functions by their special 
representations that were derived by Humblet~\cite{Humblet}. He showed  that 
the regular and irregular Coulomb functions have the following structure 
(see also Refs. \cite{Rakityansky2013,Rakityansky2022}):
\begin{eqnarray}
\label{F_tilde}  
   F_{\ell}(\eta,kr) &=& D_{\ell}(\eta,k) \tilde{F}_{\ell}(E,r)\ ,\\
   G_{\ell}(\eta,kr) &=& M(\eta) D_{\ell}(\eta,k) \tilde{F}_{\ell}(E,r)
   + 
   \frac{k}{D_{\ell}(\eta,k)}  \tilde{G}_{\ell}(E,r)\ ,  \label{G_tilde}
\end{eqnarray}
where $\tilde{F}_{\ell}(E,r)$ and $\tilde{G}_{\ell}(E,r)$ are  
single-valued and analytic in $E$. 
% These expressions 
% were used to derive Eq.~(\ref{Jost_semi-analytic}) in Refs. 
% \cite{Rakityansky2013,Rakityansky2022}. 
Substituting these expressions  in Eqs.~(\ref{A_R}) 
and (\ref{B_R}), we obtain:
\begin{eqnarray} 
        A_{\ell}(E) =  
        \frac{\tilde{G}_{\ell}(E,a)}
        {Q_{\ell}(E,B_R)}
        - 
        \left[
        a\left.
        \frac{d\tilde{G}_{\ell}(E,r)}{dr}\right|_{r=a}
        -B_R \tilde{G}_{\ell}(E,a)
        \right] \frac{R_{\ell}(E,B_R)}{Q_{\ell}(E,B_R)}
        \nonumber \\
        \label{A_R_analytic}  
\end{eqnarray}
and
\begin{eqnarray} 
        B_{\ell}(E) = -  
        \frac{\tilde{F}_{\ell}(E,a)}
        {Q_{\ell}(E,B_R)}
        + 
        \left[
       a\left.
        \frac{d\tilde{F}_{\ell}(E,r)}{dr}\right|_{r=a}
         -B_R \tilde{F}_{\ell}(E,a)
        \right] \frac{R_{\ell}(E,B_R)}{Q_{\ell}(E,B_R)}.
        \nonumber \\
        \label{B_R_analytic}
\end{eqnarray}
The dependence on the odd powers of $k$ (which causes the square-root 
branching, $k\sim\sqrt{E}$) as well as on $M(\eta)$ (which causes the 
logarithmic branching) have cancelled out and therefore $A_{\ell}(E)$ and 
$B_{\ell}(E)$ are single-valued in variable $E$.  

What about analyticity of $A_{\ell}(E)$ and $B_{\ell}(E)$ given by 
Eqs. (\ref{A_R_analytic}, \ref{B_R_analytic})? In 
order to answer this question, we re-write the representations 
(\ref{R_matrix}, \ref{Q_fitting}) as
\begin{equation}
\label{R_P}
   R_\ell(E,B_R)=
   \frac{1}{\mathscr{P}_0(E)}\sum_{n=1}^N\gamma_{n\ell}^2\mathscr{P}_n(E)\ ,
\end{equation}
\begin{equation}
\label{Q_P}
   Q_\ell(E,B_R)=
   \frac{1}{\mathscr{P}_0(E)}\sum_{n=1}^N
   \lambda_{n\ell}\gamma_{n\ell}\mathscr{P}_n(E)\ ,
\end{equation}
where $\mathscr{P}_n(E)$ ($n=0,1,2,\dots N$) are the following polynomials:
\begin{equation}
\label{P_0}
   \mathscr{P}_0(E)=\prod_{n=1}^N(E_{n\ell}-E)\ ,
\end{equation}
\begin{equation}
\label{P_n}
   \mathscr{P}_n(E)=\frac{\mathscr{P}_0(E)}{E_{n\ell}-E}\ ,
   \qquad\text{for}\ n>0\ .
\end{equation}
It is clear that both $R_{\ell}(E,B_R)$ and $Q_{\ell}(E,B_R)$ are singular at 
all $N$ points $E=E_{n\ell}$, $n=1,2,\dots,N$. However, in their ratio, 
$R_{\ell}/Q_{\ell}$, the denominators $\mathscr{P}_0(E)$ cancel out, which 
removes these singularities. The inverted $1/Q_{\ell}(E,B_R)$ is non-singular 
(equals to zero) at these poles as well. Unfortunately, these facts are not 
sufficient to make $R_{\ell}/Q_{\ell}$ and $1/Q_{\ell}$ analytic. The numerator 
on the right-hand side of Eq. (\ref{Q_P}) is a polynomial of the order $(N-1)$ 
in variable $E$ and thus has $(N-1)$ zeros in the complex $E$-plane. As a 
consequence, the right hand sides of Eqs. (\ref{A_R_analytic}, 
\ref{B_R_analytic}) have first-order poles at these zeros.

Therefore the functions $A_{\ell}(E)$ and $B_{\ell}(E)$, when obtained within 
the $R$-matrix theory, are single-valued (as it should be), but are not 
analytic on the whole $E$-plane. They are singular at $(N-1)$ isolated points, 
i.e. they are meromorphic. This fact illustrates our statement given in the 
Introduction, that the analytic structure of the $R$-matrix is obscure, which 
makes its analytic continuation to complex energies not completely reliable. 
This is where the Jost function (which is free of such defects) could be of 
help.

%%%%%%%%%%%%%%%%%%%%%%%%%%%%%%%%%%%%%%%%%%%%%%%%%%%%%%%%%%%%%%%%%%%%%%%%%%%%
\section{Modified Jost-function analysis} 
\label{section_alternative_Jost}
%%%%%%%%%%%%%%%%%%%%%%%%%%%%%%%%%%%%%%%%%%%%%%%%%%%%%%%%%%%%%%%%%%%%%%%%%%%%
The semi-analytic representations of the Jost 
functions (\ref{Jost_semi-analytic}) are exact and valid for any potential. When fitting experimental 
data, we vary the functions $A_\ell(E)$ and $B_\ell(E)$, which are the only 
unknown factors in these representations. In order to vary $A_\ell(E)$ and 
$B_\ell(E)$, we can choose any appropriate parametrization, with the 
condition that they are single-valued and analytic on the whole complex 
$E$-plane.

Originally, it was suggested to use several terms of the Taylor series to 
parametrize these functions, as is given by Eq. (\ref{orig_Taylor}).
A Taylor series gives a good approximation of a function near the central point 
of the expansion, i.e. around $E=E_0$, in our case. When $E$ moves away 
from $E_0$, the accuracy drops. This is why, within such an approach, one 
has to use many Taylor terms. Even with a large number of terms, the energy range where the fitting is good, remains narrow.

In order to circumvent this difficulty, we can use the parametrization 
inspired by the $R$-matrix theory. This can be done in the following 
way. Looking at Eqs. (\ref{A_R_analytic}, \ref{B_R_analytic}), we see that
most of the factors are the same for any physical system, i.e. they are just some 
combination of the Coulomb-related functions and do not depend on the 
potential. The only factors determined by the potential, are the 
functions $R_{\ell}$ and $Q_{\ell}$ given by Eqs. (\ref{R_P}, \ref{Q_P}).
Therefore $A_{\ell}(E)$ and $B_{\ell}(E)$ can be represented as linear 
combinations of the polynomials $\mathscr{P}_n(E)$ ($n=0,1,2,\dots,N$). As was 
discussed in the preceding section, in these equations, a common denominator is 
present, which is a polynomial of the order $(N-1)$. Since this polynomial makes 
$A_{\ell}(E)$ and $B_{\ell}(E)$ meromorphic instead of analytic and since it 
cancels out in the expression (\ref{S_AB}) for the $S$-matrix anyway, we can 
ignore it. Perhaps it should be emphasized that the choice of the 
parametrization of the functions $A_{\ell}(E)$ and $B_{\ell}(E)$ is rather 
arbitrary. We therefore have some freedom in this, and we do not violate or distort 
something by ignoring the common denominator. 

Based on this simple reasoning, we replace the Taylor expansions 
(\ref{orig_Taylor}) with the following:
\begin{eqnarray} 
     A_{\ell}(E)  = \sum_{n=0}^N\alpha_{n\ell}\mathscr{P}_n(E), \qquad
 \label{A_R-and_B_R_def}
     B_{\ell}(E) = \sum_{n=0}^N\beta_{n\ell}\mathscr{P}_n(E)\ , 
\end{eqnarray}
where the real numbers $\alpha_{n\ell}$ and $\beta_{n\ell}$ can be used as  
fitting parameters. In contrast to the Taylor series (\ref{orig_Taylor}), the 
parameters in Eqs. (\ref{A_R-and_B_R_def}) do not depend on the  
(arbitrarily chosen) central point $E_0$. This new parametrization incorporates 
the prior (approximate) knowledge of the resonance energies, $E_{n\ell}$, which 
should make the fitting of the data easier and more accurate. 

%%%%%%%%%%%%%%%%%%%%%%%%%%%%%%%%%%%%%%%%%%%%%%%%%%%%%%%%%%%%%%%%%%%%%%%%%%%%%%
\section{Fitting procedure} 
\label{section_fitting}
%%%%%%%%%%%%%%%%%%%%%%%%%%%%%%%%%%%%%%%%%%%%%%%%%%%%%%%%%%%%%%%%%%%%%%%%%%%%%%
In a typical scenario, a set of experimental total cross-section data points is available, 
\begin{equation}
\label{fit.1}
   \sigma_{\mathrm{total}}\left(E_i\right) \pm \Delta_i\ ,
   \qquad
   i=1,2,\dots,N^{(\mathrm{data})}
\end{equation}
with the corresponding experimental errors $\Delta_i$, measured at the collision 
energies $E_i$. In order to fit this set of data points, we parametrize the functions 
$A_\ell(E)$ and $B_\ell(E)$ as is given by Eqs. (\ref{A_R-and_B_R_def}). We then 
substitute them in the $S$-matrix (\ref{S_AB}) and obtain the total cross-section (\ref{sigma_total}) by summing over several partial cross-sections 
(\ref{sigma_partial}) from $\ell=0$ up to some $\ell_{\mathrm{max}}$, that we 
consider as giving a noticeable contributions to the total cross-section. As a result of 
such a summation, for each data point we obtain the fitted cross-section 
$\sigma_{\mathrm{fit}}(E_i)$, which depends on the factors 
$\alpha_{n\ell}$ and $\beta_{n\ell}$, as well as the energies, $E_{n\ell}$, that 
are a part of the polynomials (\ref{P_0}, \ref{P_n}).

The set of the energies $E_{n\ell}$ is chosen in the same way as in the 
$R$-matrix analysis. In simple words, this is done as follows: Inspecting the 
experimental data, we find the collision energies around which 
$\sigma_{\mathrm{total}}\left(E\right)$ shows some irregularities, i.e. where it is 
noticeably different from a smooth monotone function. Such points are 
considered as ``suspected resonances'' and are included in the set of 
$E_{n\ell}$. It should be emphasized that in most cases, the final real 
parts of the resonance energies (\ref{resEnergy}) differ from the chosen 
$E_{n\ell}$. This means that any small shifts (either to the left or to the 
right) of $E_{n\ell}$ can only change the optimal values of $\alpha_{n\ell}$ 
and $\beta_{n\ell}$, but do not affect the final results. However, the closer a 
chosen $E_{n\ell}$ is to the actual $E_r$, the better the fit. In a sense, 
the choice of these energy points, based on the inspection of the data, is a 
kind of prior knowledge that is embedded in the fitting procedure from the 
outset. 

The ``suspected resonances'' are not the only energy-points that constitute the 
set $E_{n\ell}$.  A good fit of the data can only be 
achieved if, in addition to them, some background energies are included. Usually they are chosen far away (either to the left or to the 
right) of the energy interval where the measurements were done. The choice of 
these background energies is made by trial and error.

In contrast to the values of $E_{n\ell}$, for the combination coefficients 
$\alpha_{n\ell}$ and $\beta_{n\ell}$ there is no ``prior knowledge'' and 
therefore they are treated as the free 
adjustable parameters. The optimal values are found by minimizing the 
following $\chi^2$ function:
\begin{equation}
\label{chi2}
   \chi^2 =
   \displaystyle
   \sum_{i=1}^{N^{(\mathrm{data})}}\left[\frac{\sigma_{\mathrm{total}}(E_i)-
   \sigma_{\mathrm{fit}}(E_i)}
   {\Delta_i}\right]^2\ .
\end{equation}
The number of adjustable parameters, $M$, depends on the number of terms in Eqs. (\ref{A_R-and_B_R_def}), $N$, as well as on $\ell_{\mathrm{max}}$,
\begin{equation}
\label{Mpar}
     M=2(N+1)(\ell_{\mathrm{max}}+1)\ .
\end{equation}

%%%%%%%%%%%%%%%%%%%%%%%%%%%%%%%%%%%%%%%%%%%%%%%%%%%%%%%%%%%%%%%%%%%%%%%%%%%%%%
\section{Numerical example} 
\label{section_example}
%%%%%%%%%%%%%%%%%%%%%%%%%%%%%%%%%%%%%%%%%%%%%%%%%%%%%%%%%%%%%%%%%%%%%%%%%%%%%%
In order to test the proposed method, we use artificial ``experimental'' data, 
i.e. the cross-section data points that were not measured but calculated using a 
given potential. The uncertainties, $\Delta_i$, for such artificial data points 
are the same for all $i$, and can therefore be ignored. The spectrum of the 
resonant states for this potential is 
known. After fitting such artificial data, the resonance energies can be found 
as zeros of the Jost function, $f_\ell^{(\mathrm{in})}(E)$, on the unphysical 
sheet of the Riemann surface. Comparing the results of this search with the known 
resonance spectrum, we can judge how efficient and accurate the proposed 
approach is.

The artificial data points were generated with the help of the well-known  
Noro-Taylor potential with a Coulomb tail (see, for example, Ref. 
\cite{Sofianos1996}),
\begin{equation} 
\label{potential}
   V(r) = \frac{\hbar^2}{2\mu} 
   \left(
   7.5r^2e^{-r}+\frac{2k\eta}{r}
   \right),
\end{equation}
with $\mu=1$, $\eta = -1/(2k)$ and $\hbar c= 1 $. The partial-wave cross-sections for this potential can be calculated using any of the many known 
methods for solving a two-body scattering problem. One of these methods is 
based on the direct calculation of the Jost functions   
(see Refs. \cite{Rakityansky2022,Sofianos1996}), which was used here to find 
the cross-sections for  $\ell = 0$, $\ell = 1$ and 
$\ell_{\mathrm{max}} = 2$. The total cross-section was then found as the 
sum of these three partial cross-sections. Some resonances for this 
potential occur for each 
$\ell$. In particular, the resonances for $\ell=0$ were found in Ref. 
\cite{Sofianos1996}.  
  
A relatively small number of $40$ data points for the total cross-section were 
generated in the range $1.7 < E < 5$. These ``experimental'' values were then fitted using Eqs.~(\ref{S_AB}), (\ref{sigma_total}), (\ref{sigma_partial}), and the \\ parametrization (\ref{A_R-and_B_R_def}), with $N=3$. Since $\ell_{\mathrm{max}}=2$, the number of fitting parameters, $\alpha_{n\ell}$ and 
$\beta_{n\ell}$, defined by Eq. (\ref{Mpar}), was $M=24$. The values for 
$E_{n\ell}$, 
apart from the background values, were chosen near known values of $E_r$, as is 
also generally done with $R$-matrix fittings. The optimal values for 
$\alpha_{n\ell}$ and $\beta_{n\ell}$ were found by minimizing the $\chi^2$ 
function (\ref{chi2}) with the help of the minimization routine MINUIT 
\cite{MINUIT}. The chosen $E_{n\ell}$ and the 
fitting parameters obtained from the analysis, are given in Table 
\ref{table_parameters}.
The quality of the fitting can be seen in Fig. 
\ref{exp_fit}. The exact and the fitted curves practically coincide 
within the energy range where the ``data'' points are available.   

The resonance energies (\ref{resEnergy}) were located by finding the zeros of 
the Jost function,
\begin{equation}
\label{FinZero}
  f_\ell^{(\mathrm{in})}(E)=0\ ,
\end{equation}
on the unphysical Riemann energy sheet. It should be 
mentioned that due to the presence of the Coulomb term in the potential, the 
Riemann surface has spiral topology with an infinite number of sheets 
\cite{Rakityansky2022}. The choice of the correct unphysical sheet is done by 
taking the appropriate sign in front of the square root $k=\pm\sqrt{2\mu E}$ and by 
choosing the principal branch of the logarithmic term in Eq. (\ref{heta}). 
These choices determine the momentum-dependent and Coulomb-related coefficients 
in the representation (\ref{Jost_semi-analytic}). Since $A_\ell(E)$ and 
$B_\ell(E)$ are single-valued, they are the same for all the sheets. Such a 
separation of all the factors in the Jost function into multivalued (which are 
always given exactly) and single-valued (which are fitted) parts, is an advantage of 
the Jost-function method, because the choice of the correct Riemann sheet is 
always guaranteed.

The exact values of the resonance energies calculated from the potential and 
those obtained via the fitting are given in Table \ref{table_resonances}. The 
values obtained from the fitting are in good agreement with the exact values. 
Despite the fact that the first and second resonances for $\ell=0$ are far 
apart from  each other and that the second resonance is relatively wide, they 
are reproduced within the same fitting.

In order to simulate a real-life situation, the fit was done for the 
total cross-section, i.e. possible deviations of the partial-wave cross-sections from their exact values were not monitored during the fit. It 
turned out, however, that individual $\sigma_\ell(E)$ were nicely reproduced as 
well. This can be seen in Figs. \ref{fit_lis0}-\ref{fit_lis2}.
Around the resonance energies, the difference of the fitted 
cross-section curves for $\sigma_\ell(E)$ from the corresponding exact curves 
is practically indiscernible. At those energies, where there are 
no resonances, the fitted cross-sections are not very accurate (see, for 
example, the discrepancies near $E=2$ for $\sigma_\ell(E)$ with $\ell=1$ and 
$\ell=2$). This is not something specific to the Jost-function method. 
Similar inaccuracies also appear in $R$-matrix fittings. 

Since the proposed approach combines the original Jost-function method  (that  
uses the Taylor series parametrization) with the $R$-matrix method, it is 
logical to compare the new method with both old ones.

For the sake of comparison of the new method with the original Jost-function 
one, we made an attempt to fit the same data for 
$\ell=0$ within the original method. It turned out that the first two $S$-wave 
resonances could not be found simultaneously from a single fitting.  Either the first or 
the second of these resonances could be recovered when the energy range was 
split into two sections (around the corresponding $E_r$). Two different fittings with a large number of fitting parameters were required. In contrast, within the modified ($R$-matrix 
type) Jost-function method, the lowest resonances for $\ell=0$, $\ell=1$ and 
$\ell=2$ could be found at the same time from a single fitting of the total 
cross-section. 

%%%%%%%%%%%%%%%%%%%%%%%%%%%%%%%%%%%%%%%%%%%%%%%%%%
\begin{table*}[]
\centering
\resizebox{0.7\textwidth}{!}{%
\begin{tabular}{llllll}
\hline
$\ell$                                & 
$n$                                   & 
$E_{n\ell}$                           & 
\multicolumn{1}{c}{type}              & 
\multicolumn{1}{c}{$\alpha_{n\ell}$}  & 
\multicolumn{1}{c}{$\beta_{n\ell}$}   
\\
\hline
\\
\multirow{4}{*}{0} 
& 0 &        &            & $\textcolor{white}{+}3.8898\times 10^5 $ & 
$\textcolor{white}{+}4.9502\times10^4$    \\
& 1 & $1.78$ & resonance  & $-19.967$                                & 
$\textcolor{white}{+}30.567$              \\
& 2 & $4.0$  & resonance  & $-2.0212\times10^5$                      & 
$\textcolor{white}{+}3.0482\times 10^4$   \\
& 3 & $0$    & background & $\textcolor{white}{+}5.3628\times 10^5$  & 
$\textcolor{white}{+}9.7332\times 10^3$   \\
\\
\multirow{4}{*}{1} 
& 0 &        &            & $\textcolor{white}{+}23.197$             & 
$-8.3005$                                 \\
& 1 & $3.85$ & resonance  & $\textcolor{white}{+}8.5161$             & 
$-5.3308$                                 \\
& 2 & $4.75$ & resonance & $\textcolor{white}{+}180.56$              & 
$-6.5849$                                 \\
& 3 & $0$    & background & $-185.34$                                & 
$-75.727$                                 \\
\\
\multirow{4}{*}{2} 
& 0 &        &           & $\textcolor{white}{+}1.3990$             & 
$\textcolor{white}{+}1.4942$              \\
& 1 & $4.9$  & resonance & $-11.081$                                 & 
$\textcolor{white}{+}0.63462$             \\
& 2 & $20.0$ & background & $\textcolor{white}{+}96.745$             & 
$-17.167$                                 \\
& 3   & $0$  & background & $\textcolor{white}{+}14.314$             & 
$-17.277$                                 \\
\hline
\end{tabular}%
}
\caption 
{
Chosen $E_{n\ell}$ and the corresponding fitting parameters obtained when
the artificial total cross-section data-points were fitted using the 
modified Jost-function method. The data points were generated for the 
potential (\ref{potential}) within the energy interval $1.7<E<5$.
The ``type'' indicates if the corresponding 
parameters are  associated with an approximately known resonance energy or if 
it is related to the background cross-section. According to Eqs. (\ref{P_0}), 
(\ref{P_n}) and (\ref{A_R-and_B_R_def}), there are no $E_{n\ell}$ for $n=0$. 
Since this is a model calculation, the units of all the values correspond to 
$\hbar c=1$.
}
\label{table_parameters}
\end{table*}
%%%%%%%%%%%%%%%%%%%%%%%%%%%%%%%%%%%%%%%%%%%%%%%%%
\begin{table*}[]
\centering
\resizebox{\textwidth}{!}{%
\begin{tabular}{cllllll}
\hline
\multicolumn{1}{l}{} & \multicolumn{2}{c}{Exact}                                & \multicolumn{2}{c}{Jost function}                                  & \multicolumn{2}{c}{$R$-matrix}                             \\
$\ell$               & \multicolumn{1}{c}{$E_r$} & \multicolumn{1}{c}{$\Gamma$} & \multicolumn{1}{c}{$E_r$} & \multicolumn{1}{c}{$\Gamma$} & \multicolumn{1}{c}{$E_r$} & \multicolumn{1}{c}{$\Gamma$}
 \\
 \hline
\multirow{3}{*}{0}   & $1.780524536$             & $9.5719\times 10^{-5}$       & $1.780524652$             & $9.6022\times 10^{-5}$        & $1.780000295$                  & $9.7589\times 10^{-5}$       \\
                     & $4.101494947$             & $1.157254423$                & $4.100074051$             & $1.126491452$                & $3.984433773$                  & $0.733390271$                \\
                     & $4.6634611068$            & $5.366401527$                & \hspace{11mm}$-$          & \hspace{11mm}$-$             & \hspace{11mm}$-$          & \hspace{11mm}$-$               \\ \\
\multirow{2}{*}{1}   & $3.8480016342$            & $0.275384458$                & $3.854146542$             & $0.272421209$                & $3.848361301$                  & $0.308781678$                    \\
                     & $4.7500534831$            & $3.505579863$                & \hspace{11mm}$-$          & \hspace{11mm}$-$             & $3.714417102$                  & $2.686407215$                     \\ \\
\multirow{2}{*}{2}   & $4.9005161451$            & $1.567507025$                & $4.802927949$             & $1.423821138$                & $4.826534238$                  & $0.544536450$                    \\
                     & $5.3006134745$            & $5.884714883$                & \hspace{11mm}$-$          & \hspace{11mm}$-$             & \hspace{11mm}$-$          & \hspace{11mm}$-$            
\\
\hline
\end{tabular}%
}
\caption 
{
Exact resonance parameters obtained from the calculations for the 
potential (\ref{potential}) and the corresponding values found via  
the modified Jost-function and $R$-matrix fit of the artificial data points for 
$\sigma_{\mathrm{total}}(E)$ in the energy interval $1.7<E<5$. The units of all 
the values correspond to $\hbar c=1$.
}
\label{table_resonances}
\end{table*}
%%%%%%%%%%%%%%%%%%%%%%%%%%%%%%%%%%%%%%%%%%%%%%%
% R-matrix
% For l=0:
% Er = 1.7800002954276719 Gamma = 9.7589308832987900E-005   
% Er = 3.9844337727074803 Gamma = 0.73339027121802969   
% 
% For l=1:
% Er = 3.8483613013235538 Gamma = 0.30878167829590636
% Er = 3.7144171017530381 Gamma = 2.6864072149649250
% 
% For l=2:
% Er = 4.8265342382841316 Gamma = 0.54453645071049150
%%%%%%%%%%%%%%%%%%%%%%%%%%%%%%%%%%%%%%%%%%%%%%%%%%%%%%%%%%%%%%%%%%%%%%%%%%%%
%%%%%%%%%%%%%%%%%%%%%%%%%%%%%%%%%%%%%%%%%%%%%%%%
\begin{figure*}[htbp]
\centerline{\includegraphics[width=\textwidth]{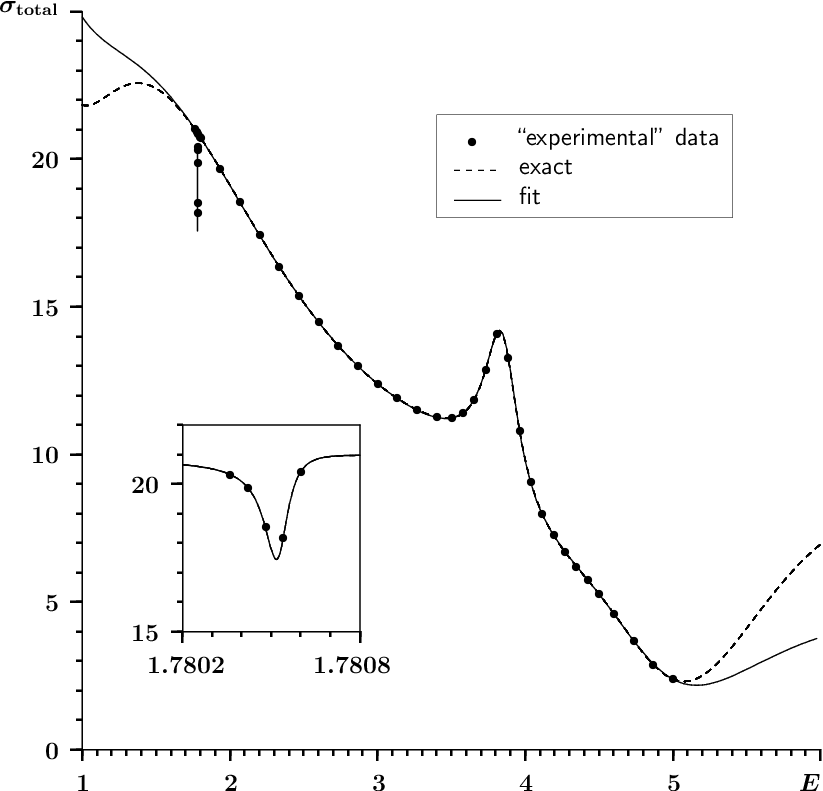}}
\caption
{
The total cross-section artificial data points (dots) generated using the  
potential (\ref{potential}). The dashed curve shows the exact cross-section, 
while the solid curve is what is obtained via the modified Jost-function fit. 
The cross-section 
$\sigma_{\mathrm{total}}(E)$ in 
the energy interval $1.7802\leqslant E\leqslant 1.7808$ around the sharp 
resonance is shown in the insert using a larger energy-scale. The exact and 
fitted curves are practically indistinguishable in this interval.
}
\label{exp_fit}
\end{figure*}
%%%%%%%%%%%%%%%%%%%%%%%%%%%%%%%%%%%%%%%%%%%%%%%%%
\begin{figure*}[htbp]
\centerline{\includegraphics[width=\textwidth]{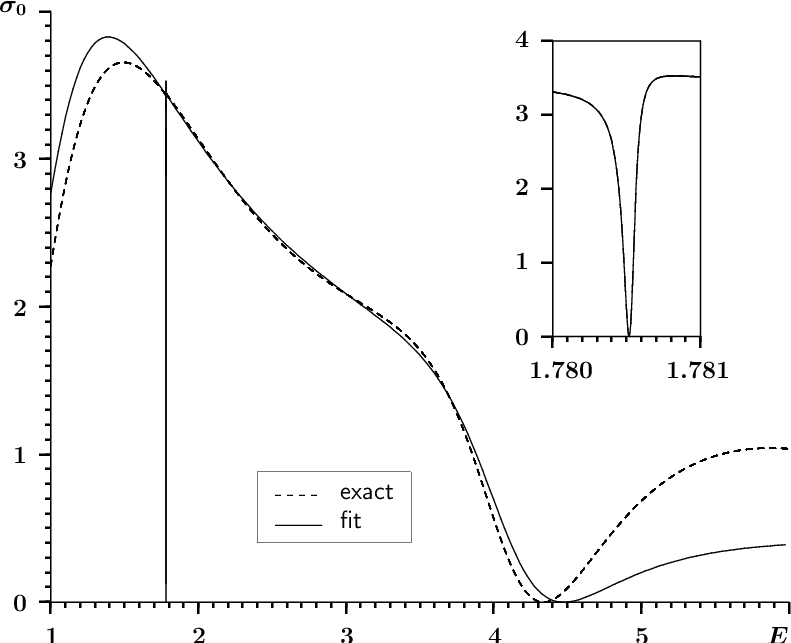}}
\caption
{
Exact (dashed curve) and fitted (solid curve) partial-wave cross-sections for 
$\ell=0$. The cross-section $\sigma_0(E)$ in 
the energy interval $1.780\leqslant E\leqslant 1.781$ around the sharp 
resonance is shown in the insert using a larger energy-scale. The exact and 
fitted curves are practically indistinguishable in this interval.
}
\label{fit_lis0}
\end{figure*}
%%%%%%%%%%%%%%%%%%%%%%%%%%%%%%%%%%%%%%%%%%%%
\begin{figure*}[htbp]
\centerline{\includegraphics[width=\textwidth]{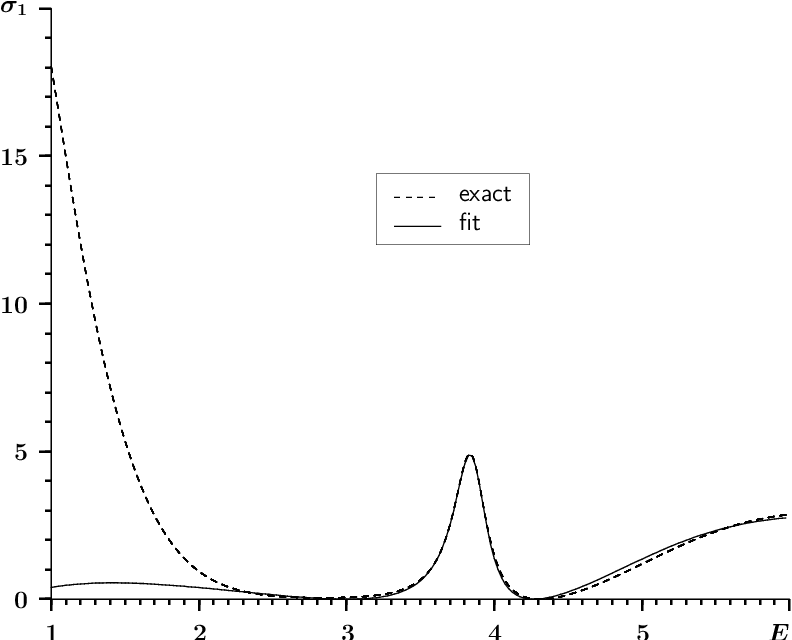}}
\caption
{
Exact (dashed curve) and fitted (solid curve) partial-wave cross-sections for 
$\ell=1$.
}
\label{fit_lis1}
\end{figure*}
%%%%%%%%%%%%%%%%%%%%%%%%%%%%%%%%%%%%%%%%%%
\begin{figure*}[htbp]
\centerline{\includegraphics[width=\textwidth]{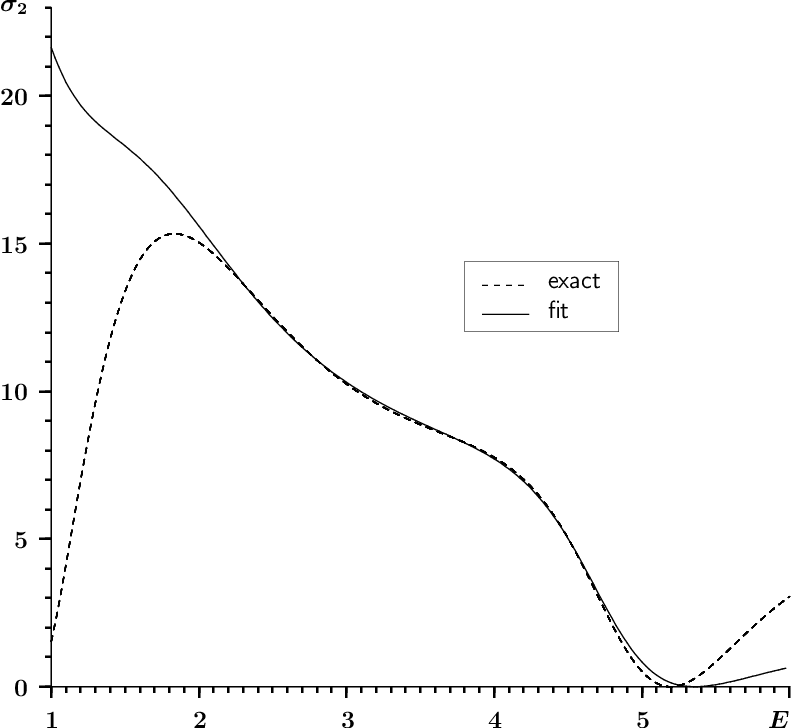}}
\caption
{
Exact (dashed curve) and fitted (solid curve) partial-wave cross-sections for 
$\ell=2$.
}
\label{fit_lis2}
\end{figure*}
%%%%%%%%%%%%%%%%%%%%%%%%%%%%%%%%%%%%%%%%

For the second comparison, we fitted the same artificial experimental data 
within  the standard $R$-matrix approach. In doing this we chose $B_R=0$ and 
tried to use the same values of $E_{n\ell}$ as in the new method. However, a reasonable fit was only obtained when significantly adjusting the $E_{n\ell}$ values that correspond to the background. Moreover, we had to choose 
different values of the channel radius for different $\ell$, namely, 
$a=0.53,\ 1.31,\ 0.94$ for $\ell=0, 1, 2$, respectively. The optimal parameters 
of the $R$-matrix thus obtained are given in Table \ref{table.R_parameters}. 
The exact and $R$-matrix-fitted total-cross-section curves are depicted in Fig. 
\ref{fig.R}. The resonances located as the poles of the corresponding 
$S$-matrix are given in Table \ref{table_resonances}. As is seen, the 
accuracy of the obtained resonance parameters is lower as compared to the 
modified Jost function method. Besides that, the $R$-matrix fit requires more 
effort in finding appropriate values of the arbitrary parameters such as the 
channel radius and the background $E_{n\ell}$.

%%%%%%%%%%%%%%%%%%%%%%%%%%%%%%%%%%%%%%%%%%%%%%%%%%
\begin{table*}[]
\centering
%\resizebox{0.7\textwidth}{!}{%
\begin{tabular}{lllll}
\hline
$\ell$                                & 
$n$                                   & 
$E_{n\ell}$                           & 
\multicolumn{1}{c}{type}              & 
\multicolumn{1}{c}{$\gamma_{n\ell}$}  
\\
\hline
\\
\multirow{4}{*}{0} 
& 1 & $0$    & background & $0.36862$ \\
& 2 & $1.78$ & resonance  & $0.59070 \times 10^{-2}$        \\
& 3 & $4.0$  & resonance  & $0.44958  $                      \\
& 4 & $20$   & background & $ 0.88622 $  \\
\\
\multirow{4}{*}{1} 
& 1 & $0$    & background & $ 1.4806 $     \\
& 2 & $3.85$ & resonance  & $ 0.19240$     \\
& 3 & $10  $ & background & $ 0.10105\times 10^{-3} $     \\
& 4 & $20$   & background & $ 3.0343 $                      \\
\\
\multirow{3}{*}{2} 
& 1 & $0$    & background & $  1.2198$    \\
& 2 & $4.9$  & resonance &  $  0.3462$                     \\
& 3 & $20.0$ & background & $  1.5640 $    \\
\hline
\end{tabular}%
%}
\caption 
{
Chosen $E_{n\ell}$ and the corresponding fitting parameters 
$\gamma_{n\ell}$ obtained when the artificial total cross-section data-points 
were fitted using the $R$-matrix (\ref{R_matrix}) with $B_R=0$. 
The data points were generated for the 
potential (\ref{potential}) within the energy interval $1.7<E<5$.
The ``type'' indicates if the corresponding 
parameters are  associated with an approximately known resonance energy or if 
it is related to the background cross-section.  
Since this is a model calculation, the units of all the values correspond to 
$\hbar c=1$.
}
\label{table.R_parameters}
\end{table*}
%%%%%%%%%%%%%%%%%%%%%%%%%%%%%%%%%%%%%%%%%%%%%%%%%
%%%%%%%%%%%%%%%%%%%%%%%%%%%%%%%%%%%%%%%%%%%%%%%%
\begin{figure*}[htbp]
\centerline{\includegraphics[width=\textwidth]{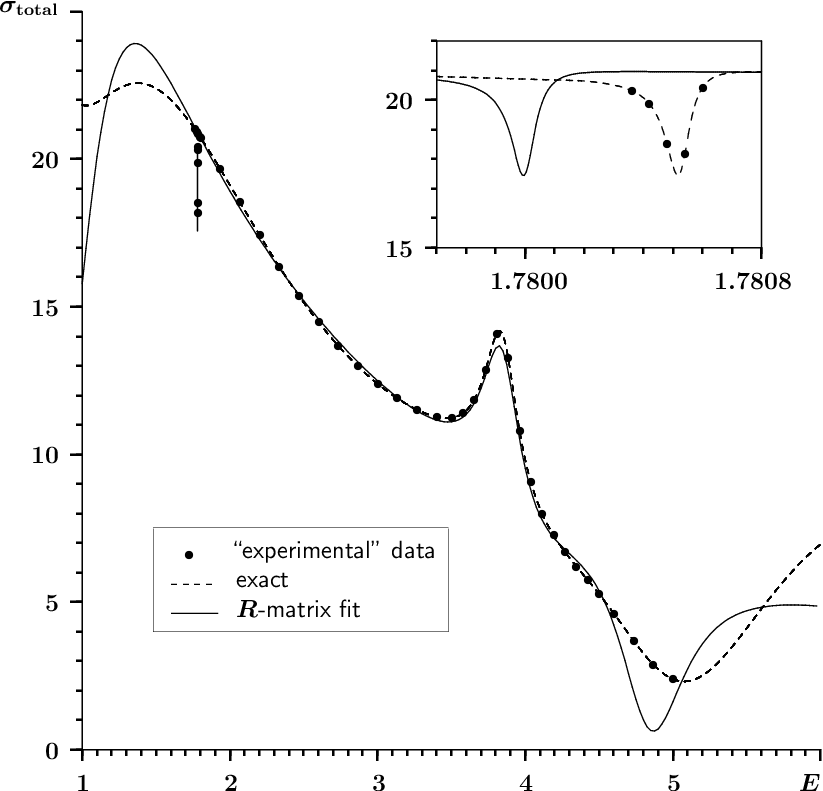}}
\caption
{
The total cross-section artificial data points (dots) generated using the  
potential (\ref{potential}). The dashed curve shows the exact cross-section, 
while the solid curve is what is obtained via the $R$-matrix fit. The cross 
section 
$\sigma_{\mathrm{total}}(E)$ in 
the energy interval $1.7797\leqslant E\leqslant 1.7808$ around the sharp 
resonance is shown in the insert using a larger energy-scale.
}
\label{fig.R}
\end{figure*}
%%%%%%%%%%%%%%%%%%%%%%%%%%%%%%%%%%%%%%%%%%%%%%%%%
\section{Conclusion}
Both the Jost-function and the $R$-matrix approaches to the analysis of 
scattering data have their specific advantages and limitations. The main goal of 
the present study was to show that a hybrid method based on the Jost function 
analysis with some elements of the $R$-matrix method, could be 
very useful in deducing the resonance parameters from experimental data.

We modified the original Jost-function method in the way the holomorphic parts 
of the Jost functions are parametrized. Instead of the Taylor series 
parametrization, we represent the holomorphic parts of the Jost functions similar to the $R$-matrix. This 
modification does not affect the analytic structure of the Jost functions. This 
allows us to correctly continue the fitted Jost function to complex energies 
and thus to reliably locate possible resonance spectral points. 

In addition to that, far fewer fitting parameters are needed to accurately fit 
the available data. This is still more than is usually required in 
the $R$-matrix fits, but here there is no dependence on arbitrary values 
such as the channel radius, $a$, and the boundary parameter, $B_R$.

Possible generalization of the hybrid method presented here, could be its 
extension to the analysis of data on the inelastic collisions with several 
binary channels. 
In the multi-channel problems the Jost function becomes a matrix and the 
Riemann surface of the energy $E$ (on which it is defined) becomes much more 
complicated, with many branch points and intricate topology 
\cite{Rakityansky2022}. In such a problem, for analytic continuation of the $S$-matrix from the real $E$-axis it is crucial to have proper analytic structure of 
the parametrization. This can be guaranteed within the Jost-matrix approach. 
However, the number of the fitting parameters in the original method (with 
the Taylor series parametrization) is too 
large and rapidly grows with the number of channels. This difficulty could 
be circumvented by incorporating some elements of the $R$-matrix theory, as is 
done here for a single-channel problem.

An example where such a multi-channel hybrid approach could be very useful, is 
the search for so called ``shadow'' poles 
\cite{Brown,PRL59,Karnakov,Bogdanova,Betan,He5}. These singularities of the $S$-matrix are located at complex $E$ within some unusual domains of the Riemann 
surface that are different from those where the resonances are. In order to 
reach these domains, the $S$-matrix with proper analytic 
structure must be used.

%%%%%%%%%%%%%%%%%%%%%%%%%%%%%%%%%%%%%%%%%%%%%%%%%%%%%%%%%%%%%%%%%%%%%%%
\section*{Acknowledgements} 
This work was funded by the South African National Research Foundation (NRF).
%%%%%%%%%%%%%%%%%%%%%%%%%%%%%%%%%%%%%%%%%%%%%%%%%%%%%%%%%%%%%%%%%%%%%%%%%%

\end{document}